\documentclass[runningheads]{llncs} % seems like 10pt
\usepackage[style=numeric-comp]{biblatex}
\bibliography{essos15}
\usepackage{alltt}
\usepackage{amstext}
\usepackage{amsmath}
\usepackage{amssymb}
\usepackage{comment}
\let\fns\footnotesize

\usepackage{mathrsfs} % for fancy A

\usepackage{graphicx}
\usepackage{type1cm}
\usepackage{eso-pic}
\usepackage{color}
\usepackage{fancyhdr}
\usepackage{floatflt}

\makeatletter
\def\thewatermark{\begin{tabular}{c}THIS SECTION\\NOT FOR PUBLICATION\end{tabular}}
\def\watermarkangle{45}
\def\watermarklightness{0.9}
\def\watermarkfontsize{2cm}
\def\watermark{%                      % starts watermarked pages
    \AddToShipoutPicture{%
            \setlength{\@tempdimb}{.5\paperwidth}%
            \setlength{\@tempdimc}{.5\paperheight}%
            \setlength{\unitlength}{1pt}%
            \put(\strip@pt\@tempdimb,\strip@pt\@tempdimc){%
        \makebox(0,0){\hss\rotatebox{\watermarkangle}{\textcolor[gray]{\watermarklightness}%
                 {\fontsize{\watermarkfontsize}{\watermarkfontsize}\selectfont{\thewatermark}}}\hss}%
        }%
    }%
}

\makeatother

\def\fract#1#2[#3]{
\frac{
\mbox{\fns$#1$}
}{
\mbox{\fns$#2$}
}
\mbox{\fns[$#3$]}
}
\def\pre[#1]#2{[#1]\mathop{{#2}}}

\def\to{{\rightarrow\kern0.5pt}}

\allowdisplaybreaks

\newenvironment{textbox}[1][]{%
  \def\@captype{box}%
    \par\nobreak\vspace{-2ex}\begin{center}\nobreak\begin{framebox}}
      {\end{framebox}\par\nobreak\end{center}\vspace{-0.5ex}}
      \newcounter{textbox}

\renewcommand{\appendix}{\par
\section*{APPENDIX -- {\em NOT FOR PUBLICATION\/}}
\setcounter{section}{0}
 \setcounter{subsection}{0}
  \def\thesection{\Alph{section}} }

\excludecomment{anonymous}\includecomment{standard}
\begin{anonymous}

\def\citeBB#1{\ifnum#1=11 \expandafter\mycite{BB11A}\else\ifnum#1=94 \expandafter\mycite{BB94A}\fi\fi}

\end{anonymous}
\begin{standard}

\def\citeBB#1{\ifnum#1=11 \expandafter\mycite{BB11}\else\ifnum#1=94 \expandafter\mycite{BB94}\fi\fi}

\end{standard}

\def\mycite#1{\cite{#1}}

\def\paragraph#1{{\bf#1}}

\begin{standard}
\def\authorA{\authorAlong}
\def\authorAemail{Peter.T.Breuer@gmail.com}
\def\authorAlong{Peter~T.~Breuer}

\def\authorB{\authorBlong\thanks{\authorBmedium{} acknowledges
the support of Museophile Limited.}}
\def\authorBemail{jonathan.bowen@bcu.ac.uk}
\def\authorBlong{Jonathan~P.~Bowen}

\def\authorBdept{School of Computing, Telecommunications and Networks}
\def\authorBinst{Birmingham City University, UK}
\end{standard}

\begin{anonymous}
\def\authorA{\authorAlong}

\def\authorAlong{Author~1}
\def\authorAemail{author 1 email}

\def\authorB{\authorBlong}

\def\authorBlong{Author~2}
\def\authorBemail{author 2 email}
\def\authorBdept{Faculty of Business}
\def\authorBinst{London South Bank University, UK}
\end{anonymous}

\title{
On the Security of Fully Homomorphic Encryption and Encrypted Computing
}
\subtitle{Is Division Safe?}

\author{
\authorA \and \authorB
}

\authorrunning{P.T.~Breuer and J.P.~Bowen}
\titlerunning{Is Division Safe?}

\institute{\authorBdept, \authorBinst\\
\email{\authorAemail}, \email{\authorBemail}
}

\date{12 September 2014}

\renewcommand{\leftmark}{\small\thepage\hspace{24pt}P.\,\,T.~Breuer and J.\,\,P.~Bowen}
\renewcommand{\rightmark}{\small Is Division Safe?\hspace{24pt}\thepage}

\begin{document}

\maketitle

\begin{abstract}
\small\em
Since fully homomorphic encryption and homomorphically encrypted
computing preserve algebraic identities such as $2*2=2+2$, a natural
question is whether this extremely utilitarian feature also sets up
cryptographic attacks that use the encrypted arithmetic
operators to generate or identify the encryptions of known
constants.  In particular, software or hardware might use encrypted
addition
and multiplication to do encrypted division and deliver the
encryption of $x/x=1$.  That can then be used to generate
$1+1=2$, etc, until a complete codebook is obtained.

\quad
This paper shows that there is no formula or computation using 32-bit
multiplication $x*y$ and three-input addition $x+y+z$ that yields a
known constant from unknown inputs.  We characterise what operations are
similarly `safe' alone or in company, and show that 32-bit division is
not safe in this sense, but there are trivial modifications that make it
so.

\end{abstract}
%\keywords{
%Fully homomorphic encryption,
%computer security,
%encrypted computing.
%}

\thispagestyle{plain}

\section{Introduction and Background}
\label{sec:Introduction}

Cryptographers have been looking for fully homomorphic encryptions since
cryptography became a modern science - Rivest (of RSA public/private key
cryptography fame) was the first to give the idea a name \cite{RSA} and
to point out that it would make it possible to carry out any kind of
operation on encrypted data without ever revealing what lies underneath.
RSA cryptography itself is partially homomorphic in that $\mbox{\it
RSA}(x,m)*\mbox{\it RSA}(y,m) = \mbox{\it RSA}(x * y,m) ~\mod~m$ for
encryption $\mbox{\it RSA}$ with modulus $m$, and that enables some
features of the digital economy today, such as being able to give change
offline from digitally encrypted money.

Thus Gentry's 2009 construction for the first time of a {\em fully}
homomorphic encryption (FHE) \cite{Gent09} -- that is, one in which
$E(x+y)=E(x)+E(y)$ as well as $E(x*y)=E(x)*E(y)$ -- was very
significant, and since then IBM in particular have devoted considerable
resources towards making the original scheme more practical.  Gentry's
encryption scheme is a realisation of Rivest's vision, in that it works
with very large integers, around the million-bit mark.  So far, the
teams working on it have got the time taken for doing a single bit
operation down to the order of a second or even less, on very powerful
vector hardware \cite{GH}, and others are employing more off-the-shelf
components (GPUs) in efforts to further commoditize the idea
\cite{Wang}, while improved schemes that may be more practicable than
the original have been proposed \cite{Coron}.  If bank accounts were
encoded in a fully homomorphic encryption uniquely known to the bank
customer, then the bank could add and subtract amounts, and add interest
to the account, without the bank ever learning what amount lies in the
account.

\paragraph{Numerical alchemy}.  The magic mentioned above is a
reflection of the fact that logical AND and (exclusive) OR are just
multiplication and addition in modulo 2 arithmetic, so an entire logic
circuit can run encrypted under a homomorphic encryption, such as the
one in an electronic calculator that computes sums and cumulative
interest.  Anything a logic circuit can do to unencrypted data, that can
also be done by combining plus and times on the encrypted data.  That
includes running a different encryption's entire decryption circuit
encrypted, which results in one being able to change the homomorphic
encryption on a piece of data without doing any decryption or
encryption.

Nevertheless, in practice, the operations on encrypted numbers that
implement a fully homomorphic encryption scheme are not quite as simple
as plain multiplication and addition -- even in the (partially
homomorphic) RSA case the remainder after dividing by $m$ has to be
taken, and fast division by a $4096$ bit number is not within easy reach
of consumer electronics.  Frequent `renormalizations' like that are
required under Gentry's and all other fully homomorphic encryption
schemes known so far (otherwise the numbers grow too big in the
intermediate calculations), and the outcome is that doing addition and
multiplication on encrypted numbers is not yet very practical.

\paragraph{The `homomorphically encrypted computing' alternative}.
The present authors showed in \cite{BB13} that it is possible to design
an entire processor in such a way that it works encrypted, provided only
that its arithmetic logic unit (ALU) satisfies certain algebraic
identities that boil down to requiring that the ALU electronics supplies
operations on encrypted numbers that correspond to plus
and times, etc, on unencrypted numbers.  The data passing through
memory, registers and buses is always in encrypted form.  There are certain
restrictions on the kind of programs that can run encrypted, because
multiple encryptions of the same values lead to {\em hardware aliasing}
\cite{BB12} and because encrypted data addresses and unencrypted program
addresses must be kept apart \cite{BB11}, but that is all.
This kind of setup
may be called {\em homomorphically encrypted computing} (HEC), and the
encrypted processor that works in this way may be called a general
purpose {\em crypto-processor unit} (KPU).

If the encryption in a KPU were a fully homomorphic encryption, then the
ALU would just implement ordinary but very long word-length computer
plus and times, with renormalizations. But there is no need to use a
fully homomorphic encryption -- any one of block-size comparable to the
processor word-size will do in principle.  The ALU electronics is
designed instead to implement whatever operations $+'$ and $*'$ are
required to give $E(x+y)=E(x)+'E(y)$ and $E(x*y)=E(x)*'E(y)$ with
respect to the encryption $E$ (note that these equations are prescriptive
for $+'$ and $*'$).  The word-size is typically close to the
conventional word-size, and there is no real need for $+'$ and $*'$ to be
as simple as ordinary plus and times.  Indeed, some secrets of the
encryption may be embedded in the hardware, because SmartCard-like
techniques \cite{SmartCard} can be used in the processor in order to
protect that data and the processes that manipulate it from physical
probes.

In contrast, the arithmetic
operators (plus and times, etc) in a fully homomorphic encryption are
implemented in software, thus open (in principle) to scrutiny, and so they
may not embed within any secret of the encryption.

\paragraph{A common attack mode}.
Whatever their relative merits, both software (FHE) and hardware (HEC)
approaches are subject to identical attacks via the natural algebra of
arithmetic.  Everyone knows that $2+2=2*2$, so what if a malicious
observer sees the encrypted arithmetic produce both $53=42+42$ and
$53=42*42$?  They should conclude that $42$ is the encryption of the
number $2$, and $53$ is the encryption of the number $4$.

In practice, an observation like that will be rare (but not never, and
once is enough).  Real encryptions use padding under the encryption that
makes it unlikely, because that induces many different encodings of each
unencrypted number, so the encryption for $4$ from $2+2$ will not be the
same as the encryption for $4$ from $2*2$.  The design for a KPU at
\url{sf.net/projects/kpu} uses 32 bits of padding for 32 bits of data in
64 physical bits, and IBM's million-bit implementation of 1-bit logic
must have an effective 999999 bits of padding.  One should expect
$4(2^{32})^3 2^{32} = 2^{130}$ computations of $2+2$ and $2*2$ in a KPU,
under the same encryption, before $2+2=2*2$ can be recognised.  Still,
there are many arithmetic identities to look out for, and each step of a
computation that an attacker can observe is one more opportunity.

The odds tilt towards an attacker, however, when the attacker can choose
the computations.  If the attacker can try $x+x$ and $x*x$ for many
(encrypted) values $x$, $x+x=x*x$ may be found allowing the attacker to
deduce what $x$ the encryption of $2$ is (the situation is more
complicated if an `ABC typing' \cite{BB14} is embedded in the ALU, which
causes $x$~op~$x$ to always give a nonsense result, for any arithmetic
operation, at the cost of trebling the size of the cipher-space).  An
attacker can choose the computation in fully homomorphic encryption,
because the arithmetic operations are precisely what are handed out in
order that computations on encrypted data may be carried out without
decryption, so an attacker can combine them into any formula that
suits.  And an attacker can also choose the sequence of instructions
to be carried out if they physically possess a KPU.  (One proviso here
is that the KPU may use security modules that reliably boot a secure
kernel that in turn only permits `officially sanctioned' codes to run,
but physical possession permits many degrees of interference with even
such a setup).  What if an attacker has a clever formula or computer
program that uses the operations on encrypted numbers to deliver the
encryption of a known constant, like 1, or 2, or 4?  With physical
possession comes the presumption that they can do that.

Indeed, if encrypted subtraction is one of the operations available,
then $x-x$ delivers the encryption of $0$ straight away, whatever value
$x$ is chosen.  If multiplication is available, then, in 32-bit
arithmetic, multiplying $x$ by itself 32 times gives $x^{2^{32}}=1$ or $0$,
depending as the $x$ chosen is odd or even.  And of course, if division
is available, then $x/x = 1$ so long as the $x$ chosen is non-zero.  All
these options allow an attacker to produce the encryption of $1$, then
$2=1+1$, then $3=2+1$, until an entire codebook of the encryption is
prepared.  At the very least that allows an attacker to modify data in a
controlled way, and may allow for the decryption of data that is already
encrypted, if it lies in that codebook.

So the situation is quite confused as to whether fully homomorphic
encryption and homomorphically encrypted computing are perhaps much more
vulnerable to cryptographic attack than might naively be expected.  This
paper is aimed at clarifying the status. It shows that multiplication
and three-input addition can never be used to construct a known constant
from unknown inputs, and characterises those operations that share that
property with them. These operations may in a certain sense be `safely'
distributed with a fully homomorphic encryption or set physically
into a KPU's ALU.

\section{A formal safety criterion and guarantee}.
In the first place, we will show that if one only supplies multiplication
$x*y$ and {\em addition-with-carry-in} $x+y+z$ (also known as {\em
three-input addition}, or {\em double-addition}) as available
operations on the encrypted data, then {\em there is no formula}
(indeed, no computation) {\em using these operations alone that
constructs a known constant from unknown inputs}.

That implies that a script-kiddie cannot plug in some arbitrary encrypted
values he/she has seen passing by into a a pre-supplied formula, execute
it in encrypted form using the FHE operations or the KPU, and have the
encryption of a known constant pop out as a result. The attacker can
then use the constant to construct a codebook.  In other words, 
multiplication and three-input addition on encrypted data are not
advantageous to a script-kiddie.

Restricting to just these two is not a perfect panacea, 
because as remarked above, repeated self-multiplication reliably
eventually constructs either 1 or 0 (however, embedding a typing scheme
in the arithmetic like the ABC scheme of \cite{BB14} sabotages that
particular attack).  So the result should be seen as a formal guarantee on
which other guarantees may be founded.

Note that there is no harm to functionality in electing to use
three-input addition instead of the conventional two-input addition,
because three-input addition is the form implemented within a standard
processor's ALU and so using it does not restrict the 
possible computations.

%Moreover, the result provides a formalisation of a limited
%invulnerability to algebraic attacks that has significance in
%the design of fully homomorphic encryptions and KPUs:
%
\begin{definition}
An operation or set of operations is said to be {\em safe} in this
context if there is no formula or computation using it or them alone
that yields a known constant from unknown inputs.
\label{def1}
\end{definition}
Multiplication and
three-input-addition are `safe' in this sense.

\section{A characterisation}.

We can characterise precisely which other
arithmetic operations play safe in the sense of Defn.~\ref{def1}
together with multiplication and three-input addition on
32 bits, and thus may be deemed suitable candidates to be distributed
along with fully
homomorphically encrypted data, or implemented in a KPU's ALU.
It turns out that they are those operations that
\begin{itemize}
\item[i.] produce zero from inputs that are all zero;
\item[ii.] produce an odd-number output from inputs that are all odd numbers.
\end{itemize}
If those conditions are violated then there is a way of constructing a
known constant in combination with multiplication and three-input
addition. If those conditions hold, then the operations are formally
safe as per Defn.~\ref{def1}, both individually and in concert with other
operations that satisfy the
same conditions, including multiplication and three-input addition.
% say f(0,0)=k !=0
% do f(g(x),g(x)) where g is a fn that takes even->0 and odd->1, g(x)=x^2^32
% that provides f(odd,odd)=j, f(even,even)=k
% if j, k are both even, then apply h(x)=x^32 to them, giving 0, constant.
% if j,k are both odd, then apply x^2^32, giving 1, constant.
% if one is odd and the other is even ... multiplying all the results
% together for all values of x gives 0 (at least 2^31 factors of 2).

\paragraph{Deciding on the safety of an operator, or fixing it to be
safe}.
In consequence, it is easy to decide if an operation $f(x,y)$ is safe, or
to alter it so it becomes safe.  Given that it is zero at
zero, it suffices to change the output by 1 at the $(x,y)$ points where
$x$ and $y$ are both odd, but $f(x,y)$ is even.

In particular one can say that division, if present in the classical
form $x/x=1$ for nonzero $x$, and $0/0=0$ say, is not safe by virtue of
$1/3=0$. That is the answer to the question in the title of this paper.

One can fix it by letting it produce the classical output $x/y$
almost always, but $1+x/y$ when $x$ and $y$ are odd but $x/y$ is even.
Multiplying the quotient by the divisor allows a program to check whether
the correction has been applied according to whether it is in the range
$(x-y,x]$ or not. It is `safe' to perform that check because \dots

\paragraph{Arbitrary patchworks of safe operations are safe}.
Curiously, a choice between two `safe' operations based on any condition
at all, whether the test is itself safe or not, is safe.  That is,
$(f(x,y)\ne0)?g(x,y):h(x,y)$ is safe if $g$ and $h$ are themselves safe
operations, whether or not the test $f$ is safe. That follows from the
characterisation.

That implies that a KPU may run instructions that implement safe
arithmetic operations during linear segments of the program, but branch
between them based on arbitrary and possibly unsafe test conditions.
The program will still implement a safe operation from the program
inputs to the program outputs, overall.  We conclude that it
is `safe' for a KPU design to offer any of the standard branch tests to
a programmer, such as comparisons $x<y$, $x=y$, etc, without regard to
whether or not the individual tests are safe in the sense of
Defn.~\ref{def1}. That is very significant in terms of the KPU's 
instruction set design, and at first sight almost unbelievable, because
the less-than operation, for example, allows the largest
integer to be identified as that (encrypted) $x$ which does not satisfy
(encrypted) $x<y$ for any (encrypted) $y$.

The correct intuition is that the test result itself is not exposed,
just the result of computation down one branch as opposed to another.
Can an attacker see the branch?  Yes, but in a KPU the instruction
opcode is encoded with respect to the standard, so the attacker does not
know which branch denotes which result of the comparison, or indeed
which comparison was made.  At any rate, the formal conclusion is that
whatever program an attacker may run in a KPU whose linear instructions
access only `safe' arithmetic and whose branches are based on arbitrary
tests, it is not guaranteed to deliver a known constant from unknown
inputs.

Despite that formal conclusion, however, one is still a long way from a
comfortable position here, because a formula that delivers say 0 half
the times and 1 the other half of the times may be formally `safe', but
statistically 50\% of the attacks based on the assumption that the
answer is 1 succeed.  Indeed, self-multiplying a random input $2^{32}$
times gives just that pattern.  So this notion of `safe' is not
sufficient on its own; it is just a minimal formal guarantee that things
are not so very extremely bad that an attacker can be absolutely sure
they have walked away with the right result.

One might think that entropy-based measures of safety would be more in
line with the statistical view of attack and defence, but if one
considers the standard two-input addition table on 1-bit of data (this
is binary XOR), then entropy measures say that there is one bit
of variation in the output, while the canny attacker will observe that
it suffices to provide identical inputs $x$ and $y$ in order to
guarantee that the output is always zero, with no variation at all.

Restricting to just multiplication and three-input addition as the operations
that it is safe to use, however, brings us to a question of coverage
that we do not presently know the answer to: what operations may be
implemented using just these two?

Experiments show that in 2-bit arithmetic, just 1282 operations are
available of the 4096 that might be formed using multiplication and
two-input addition.\footnote{The combinations of 2-bit multiplication and
(two-input) addition are characterised by (i), (ii) above and
also (iii) even inputs produce an even result, and (iv) the parity of
$f(x,y)$, $f(x+2,y)$, $f(x,y+2)$, $f(x+2,y+2)$ is always the same,
stepping a distance 2 up or down or left or right in the $x$-$y$ table
of the operator's arithmetic, and (v) the differences along opposite
edges of a 2x2 square in the operator table are always the same, in that
$f(x,y+2)-f(x,y) = f(x+2,y+2)-f(x+2,y)$ and $f(x+2,y)-f(x,y) =
f(x+2,y+2)-f(x,y+2)$.}  Complementarily, we also do not know precisely
which extra `safe' operations must be added to the set in order to be
able to form via combinations the full set of all the `safe' operators
that satisfy (i), and (ii) (there are $2^4 4^{11} = 2^{26} = 64$M of these),
or some characterisable subset such as those
which also satisfy (iii) even inputs produce an even output (there are
$2^4 2^4 4^7 = 2^{22} = 4$M of these).  The
answers to these questions also bear on the design of an encrypted ALU
in a KPU, or on which operations should be made available in public to
users of a fully homomorphic encryption.

\section{An easy argument in mod 2 arithmetic}

We will start on backing up the technical claims with an argument that
shows:

\begin{proposition}
Multiplication and three-input addition on 32-bit arithmetic are jointly
safe in  the sense of Defn.~\ref{def1}.
\end{proposition}
That is, there is no formula in
these operations that delivers a constant from unknown inputs, such as
$x/x = 1$ might produce.

We do  calculation in mod 2 arithmetic because if 32-bit
multiplication and addition can be combined into a formula that gives a
32-bit constant, then looking at everything mod 2, the same formula in
the same operations mod 2 gives the value 1 (if the constant is odd) or
0 (if the constant is even).  Either way, the result is a constant mod
2.

So, if it is proved that it is impossible to produce a constant from
these operations mod 2, which is 1-bit arithmetic, it has
been proved that it is impossible to produce a constant in 32-bit
arithmetic, which is what is wanted.  But the argument in mod 2
arithmetic is very easy:

\begin{enumerate}
\item Multiplication takes odd numbers to odd numbers. Similarly, adding up
three odd numbers gives an odd number.
\item So any formula using only multiplication and three-input addition takes
all odd inputs to odd
outputs. I.e., set all the inputs to 1 mod 2, and the output is 1 mod 2.
\item So the supposed constant, if it exists, must be 1 mod 2.
\label{step3}
\item But multiplication also takes inputs that are all even (i.e. 0 mod 2)
to an even output (i.e., 0 mod 2), and so does addition.
\item So the formula that supposedly produces a constant takes inputs that
are all 0 mod 2 to 0 mod 2.
\item That means the constant must be 0 mod 2.
\label{step6}
\item That is a contradiction (between \ref{step3} and \ref{step6}), so
the formula cannot exist.
\end{enumerate}
Three-input addition is essential in that argument, because it preserves
odd parity. Two-input addition does not do that: an odd number plus an
odd number is even.

\section{A general argument}

The following elements allow the argument in the previous section
to succeed. Operations must
\begin{itemize}
\item[i.] produce zero from inputs that are zero;
\item[ii.] produce an odd-number output from inputs that are odd numbers.
\end{itemize}
That means that there are two (disjoint) sets, \{0\} and
\{odd numbers\}, that are stable under these operations. Any other
operation that also stabilises those sets will combine arbitrarily with
other such operators, possibly multiplication and three-input addition,
to produce another operator with the same properties. Because odd
numbers are all different from zero, the formula that results gives
different values on the two sets, and in consequence is not constant.

That is a sufficiency argument for (i) and (ii). There is also a fairly
simple argument
that shows that any operator that violates condition (i) can be used as
part of a formula that 
manufactures a constant with the help of multiplication and three-input
addition, and we will elaborate it to apply to condition (ii)
too, showing that the conditions are both necessary. Here is the argument for
the necessity of condition (i):
\begin{enumerate}
\item Say that $f(0,0)=k_0$ for some $k_0 \ne 0$;
\item let $g(x)$ be the function that results from repeated
self-multiplication, $g(x)=x^{2^{32}}$. Then $g(x)$ is 0 for even $x$
and 1 for odd $x$;
\item compose $h(x)=f(g(x),g(x))$, which has the property that $h$
applied to even numbers is the $k_0 \ne 0$, and $h$ applied to odd numbers
is some $k_1=f(1,1)$ that one may as well take to be different from
$k_0$, or one has produced a constant $h(x)=k_0=k_1$ already;
\item if both $k_0$, $k_1$ are odd (we can calculate the value offline
from this proof), consider applying $g$ to them, producing the
constant $1 = g(h(x))$;
\item if at least one of $k_0$, $k_1$ are even (but we do not know
which), multiply together all the values of $h(x)$ obtained as $x$
varies through all possible values, and apply
$g$ to the result, which must be even as $k_0k_1$ is a factor,
producing the constant $0$.
\end{enumerate}
%
%So condition (i) is necessary if it is not possible to produce a
%constant result from unknown inputs.
%
For the necessity of condition (ii), consider the action of operators
$f(x,y)$ on functions $p(x)$ by substitution: $p\mapsto q$ where
$q(x)=f(p(x),p(x))$. The idea of the proof is to show first that $f$
must take a function that takes odd numbers to odd numbers to another
function that takes odd numbers to odd numbers, or else one can
construct a constant. Then one can deduce fairly immediately that the
operator $f$ must itself take odd numbers to odd numbers.

Here is the first part of the argument, showing that $f$ must stabilise
the odd-preserving functions if one cannot construct a known constant from
unknown inputs using it:
\begin{enumerate}
\item Suppose for contradiction that $f(p(x),p(x))=q(x)$
where $p$ preserves odd numbers but
$q$ does not. Then $q(x_1)=x_0$, where $x_1$ is odd and $x_0$ is even;
\item once again, apply the function $g(x)=x^{2^32}$ to $x_0$, producing
$0=g(q(x_1))$;
\item since the constant $x_1$ is odd, one can produce the function
$h(x)=x_1*x$ by repeated three-input self-addition $((x+x+x)+x+x)+x+x
\dots$, and $h(1)=x_1$. Applying $h$ first,
$0=g(q(h(1)))$ and $g(q(h(x)))$ is a
function that turns 1 into 0;
\item precede $g(q(h(x)))$  by $g(x)=x^{2^{32}}$, which
turns odd numbers into 1 and even numbers into 0, and so $g(q(h(g(x))))$
turns odd numbers into 0;
\item it also turns even numbers into some constant $g(k_0)$ where
$k_0=q(0)=f(p(0),p(0))$, since $g$ applied to an even number is 0 and $h(0)$
is 0. If $k_0$ is even, $g(k_0)=0$ and $g(q(h(g(x))))$ is a constant
function with result 0. If $k_0$ is odd, then $g(k_0)=1$ and
$g(q(h(g(x))))$ turns odd numbers into 0 and even numbers (including 0)
into 1;
\item in the latter case multiply by $g(x) = x^{2^{32}}$ to produce
$g(q(h(g(x))))*g(x) = 0$, whatever the input $x$;
\item by contradiction, then, if one cannot make a constant from unknown
inputs using the operator $f$ (in company with multiplication and
three-input addition), it must preserve the odd-preserving functions.
\end{enumerate}
Now for the second part of the argument:
preserving odd numbers is the same as preserving the
odd-preserving functions.

In one direction, if an operator $f(x,y)$ preserves odd numbers, then it
takes $p(x)$ that turns odd numbers to odd numbers to
$q(x)=f(p(x),p(x))$ that also turns odd numbers into odd numbers,
substituting through with odd $x$.

For the converse direction, if an operator preserves the odd-preserving
functions, does it necessarily preserve odd numbers?  Suppose for
contradiction that $f(x_1,y_1)=z_0$ with $x_1$, $y_1$ odd and $z_0$
even, and let $n$ be the odd number such that $y_1 = n x_1$ mod
$2^{32}$.  Then $p_2(x)= n x$ can be implemented using $(n-1)/2$
three-input additions $((x+x+x)+x+x)+x+x \dots$ and $p_2(x_1)=y_1$ and
$p_2$ is an odd-preserving function.  Apply $f$ to $p_1(x) = x$ and
$p_2$ and one gets $g(x)=f(p_1(x),p_2(x))=f(x,nx)$ which is by
hypothesis a function that takes odd numbers to odd numbers, since $p_1$
and $p_2$ both take odd numbers to odd numbers.  Then
$g(x_1)=f(x_1,x_2)=z_0$, but $x_1$ is odd and $z_0$ is even, not odd.
So $g$ is not an odd-preserving function after all, in contradiction to
the hypothesis.  So, yes, $f$ preserves odd numbers.  Thus, as claimed
earlier:
\begin{proposition}
An operator is safe in the sense of Defn.~\ref{def1}
in conjunction with multiplication and three-input addition iff it (i)
takes zero inputs to zero, and (ii) takes odd inputs to an odd output.
\end{proposition}
That allows decisions as to which operators may distributed along with a
fully homomorphic encryption and which operations may be implemented in
the ALU of a KPU or how they ought to be modified, to be taken in a
technical framework.

\section{Conclusion}
We have defined a formal notion of safety for operations made available
as part of a fully homomorphic encryption, or supplied by the ALU within
the KPU in a homomorphically encrypted computing context.  Conforming
implementations do not permit a script-kiddie to walk away with the
encryption of a known constant from applying a formulaic combination of
the operators to arbitrary unknown encrypted values that have been
observed.

We have characterised the operations that are safe in combination with
multiplication and three-input addition on 32-bit arithmetic as those
which take zero to zero and odds to odds. Every operation is at most 1
away in uniform norm from a safe variant, and the characterisation tells
one how to change it to be safe (division $x/y$ is not safe, but there
is a safe variant, to which a check can
safely be applied to tell if the correction has been made).

\printbibliography

\end{document}